\documentstyle[epsf,12pt]{article}
\newcommand{\eps}{\varepsilon}

\begin{document}

\begin{titlepage}
\begin{flushright}
IFUP--TH 2004/35 \\
\end{flushright}
~


\begin{center}
\Large\bf
Semiclassical and quantum Liouville theory on the sphere
\footnote{This work is  supported in part
by M.I.U.R.}
\end{center}

\vskip 1truecm
\begin{center}
{Pietro Menotti} \\ 
{\small\it Dipartimento di Fisica dell'Universit{\`a}, Pisa 56100, 
Italy and}\\
{\small\it INFN, Sezione di Pisa}\\
{\small\it e-mail: menotti@df.unipi.it}\\
\end{center}
\begin{center}
{Gabriele Vajente} \\  
{\small\it Scuola Normale Superiore, Pisa 56100, Italy and}\\
{\small\it INFN, Sezione di Pisa}\\
{\small\it e-mail: g.vajente@sns.it}\\
\end{center}

\vskip 0.5truecm



\vskip 0.5truecm

\begin{abstract}

We solve the Riemann-Hilbert problem on the sphere topology for three
singularities of finite strength and a fourth one infinitesimal, by
determining perturbatively the Poincar\'e accessory parameters. In
this way we compute the semiclassical four point vertex function with
three finite charges and a fourth infinitesimal. Some of the results
are extended to the case of $n$ finite charges and $m$ infinitesimal.
With the same technique we compute the exact Green function on the
sphere with three finite singularities. Turning to the full quantum
problem we address the calculation of the quantum determinant on the
background of three finite charges and the further perturbative
corrections. The zeta function technique provides a theory 
which is not invariant under local conformal transformations. Instead
by employing a regularization suggested in the case of the
pseudosphere by Zamolodchikov and Zamolodchikov we obtain the correct
quantum conformal dimensions from the one loop calculation and we show
explicitly that the two loop corrections do not change such
dimensions. We expect such a result to hold to all order perturbation
theory.
    
\end{abstract}

\end{titlepage}

\section{Introduction}

The quantization of Liouville field theory has been subject to
intensive study.
The hamiltonian quantization on a compact space, i.e. the circle, has
been carried through in \cite{CT,JW}; the main
result is that provided one properly tunes the anomalous contribution
to the energy momentum tensor, the regularized theory satisfies the
Virasoro algebra.

From the functional point of view  in the euclidean formulation a major
difficulty is that in 
absence of sources there is no stable background with sphere topology.
The situation is different in the case of the pseudosphere topology
\cite{ZZps} where a stable background solution exists around which a
perturbative expansion can be developed
\cite{ZZps,MT1,MT2}. Divergences occur in the calculation of the
graphs.  In \cite{ZZps,MT1,MT2} it was proven that 
provided 
one chooses a proper regularization of the Green function at coincident
points one reproduces the dimensions of the vertex
operators and in particular of the cosmological term as derived in
the hamiltonian approach in \cite{CT,JW}.

On the sphere, on the other hand, it is well known that in order to
have a stable background, sources have to be present, and such sources
cannot be arbitrarily small as they are subject to the Picard
inequalities. Such inequalities in particular tell us that the sources
must be at least three. On the other hand the classical
solution in presence of three singularities is known in terms of
hypergeometric functions \cite{BG}. 

The purpose of the present paper is to develop a perturbation theory
starting from the three point background such as to render the
functional approach to Liouville theory self consistent and completely
independent of the hamiltonian approach. To such end it is necessary
to compute the Green function on the classical background of three
generic charges. 

This is done in Sect.3 where the problem of computing the
semiclassical four point function with three finite charges plus a
fourth infinitesimal is addressed. The result is obtained by solving
perturbatively the Riemann-Hilbert problem, i.e. providing a
perturbative calculation of the Poincar\'e accessory parameters which
appears in the four singularity problem. The result is obtained in
terms of quadratures and as a by product the exact Green function on
the sphere with three arbitrary singularities is obtained, provided
such singularities satisfy the necessary Picard bounds.

Several of the obtained results can be extended to the
semiclassical $n+m$ point 
functions where $n$ charges are finite and the remaining $m$ 
small and this is done in Sect.4.

In order to produce the quantum correlation functions one has to
proceed with the perturbative expansion where as in \cite{ZZps,MT1,MT2} 
the expansion parameter is the Liouville coupling constant
$b$.

The first problem is to compute the functional determinant of the
linearized problem on the classical background.  The problem is
addressed in Sect.6. The most natural
tool in defining the functional determinant is the well known zeta
function regularization. However such 
regularization
gives the cosmological term
weights which are in contrast with those required by local conformal
invariance with the result that only the global $SL(2,C)$ invariance
survives.

Thus a different regularization is developed for computing such a
determinant. The procedure, which is similar in spirit to the
procedure of \cite{alvarez} is to compute the variation of the
determinant under small variation of the sources; the advantage of
such a procedure is to expose the role of the value of the Green
function at coincident points. If the Green function at coincident
points is regularized according the proposal of Zamolodchikov and
Zamolodchikov  in the context of the pseudosphere \cite{ZZps}, one
obtains dimensions for the vertex  
operators which agree with those found in the hamiltonian approach and
in particular the cosmological term assumes weight $(1,1)$.  The next
problem is to examine the higher order correction. In Sect.7 we
perform a detailed computation of the two loop contributions and we
find that using the same regularization such
contributions do not alter the quantum anomalous dimensions found in
the hamiltonian approach. In providing the necessary cancellations the
boundary terms appearing in the action play a fundamental role.

In principle such computation can be carried on to arbitrary order
perturbation theory and one could use it to furnish a perturbative
analysis of the three-point vertex function conjectured in
\cite{ZZsphere,dornotto} and derived in \cite{teschner} and of higher
point functions. Such investigation will be the subject of a future paper.

\section{Classical Liouville theory} 
 
The regularized classical action for the Liouville theory on the
sphere in presence of $N$ sources is given by
\cite{ZZsphere,takhtajan} 
\begin{eqnarray}\label{regularizedaction}
S_L[\phi] = \lim_{\stackrel{\eps_n \rightarrow 0}{R \rightarrow
\infty}} & & \Bigg\{ \int_{\Gamma_{\eps,R}} \left[ \frac{1}{\pi}
\partial_{z} \phi \partial_{\bar{z}} \phi + \mu e^{2b\phi} \right] \;
i \frac{dz \wedge d\bar{z}}{2} \nonumber \\ & & +\frac{Q}{2\pi i}
\oint_{\partial\Gamma_{R}} \phi \, \left( \frac{dz}{z} -
\frac{d\bar{z}}{\bar{z}} \right) + Q^2 \, \log R^2 \nonumber \\ & &
-\frac{1}{2\pi i} \sum_{n=1}^{N} \alpha_n \,
\oint_{\partial\Gamma_{n}} \phi \, \left( \frac{dz}{z-z_n} -
\frac{d\bar{z}}{\bar{z}-\bar{z}_n} \right) - \sum_{n=1}^{N} \alpha_n^2
\, \log \eps_n^2 \Bigg\}
\end{eqnarray}
where $z_n$ and $\alpha_n$ are position and charge of the $n$-th
source. The domain of integration is the region $\Gamma_{\eps,R} =
\left\{ |z| < R\right\} \setminus \bigcup_{n} \left\{ |z-z_n|< \eps_n
\right\}$, $\partial\Gamma_R$ is the border around infinity while
$\partial \Gamma_{n}$ is the border around the $n$-th source. Here $Q$
is a parameter linked to the transformation law of the Liouville
field. Classically its value is $\displaystyle{Q=\frac{1}{b}}$.

In order to examine the semiclassical limit of the $N$-point function
it is useful \cite{ZZsphere} to go over to the field $\varphi = 2b
\phi$. The corresponding charges are $\eta_n = \alpha_n\,b$ and the
action takes the form 

\begin{eqnarray} \label{eq:regularized-action-rescaled}
S[\varphi] = b^2 S_L[\phi] & = & \int_{\Gamma_{\eps,R}} \left[
			\frac{1}{4 \pi} \partial_{z} \varphi
			\partial_{\bar{z}} \varphi + b^2 \mu e^{\varphi}
			\right] \; i \frac{dz \wedge d\bar{z}}{2}
			\nonumber \\ 
			& &+ \frac{bQ}{4\pi i}
			\oint_{\partial\Gamma_{R}} \varphi \, \left(
			\frac{dz}{z} - \frac{d\bar{z}}{\bar{z}}
			\right) + (bQ)^2 \, \log R^2 \nonumber \\ 
			& & -\frac{1}{4\pi i} \sum_{n=1}^{N} \eta_n \,
			\oint_{\partial\Gamma_{n}} \varphi \, \left(
			\frac{dz}{z-z_n} -
			\frac{d\bar{z}}{\bar{z}-\bar{z}_n} \right) -
			\sum_{n=1}^{N} \eta_n^2 \, \log \eps_n^2.
\end{eqnarray}
The field $\varphi$ behaves like
\begin{equation}\label{eq:asymptotics}
\left\{
\begin{array}{l}
\varphi(z) = - 2\eta_n \, \log |z-z_n|^2 + O(1) \qquad {\rm for }~
z \rightarrow z_n \\ 
\varphi(z) = - 2b Q \, \log |z|^2 + O(1)
\qquad \qquad {\rm for }~ z \rightarrow \infty.
\end{array}
\right.
\end{equation}
We decompose the field $\varphi$ into the sum of a classical background
$\varphi_B$ and a quantum fluctuation $\psi = 2b\chi$, $\varphi =
\varphi_B + 2b\chi$. The action
$S_L$ becomes
\begin{equation}
S_L[\varphi_B,\chi] = S_{cl}[\varphi_B]+ S_Q[\varphi_B,\chi]
\end{equation}
where
\begin{eqnarray}\label{classicalaction}
S_{cl}[\varphi_B] & = & \frac{1}{b^2}\left[
\frac{1}{8\pi}\int_\Gamma \left(\frac{1}{2}(\partial_a\varphi_B)^2
+8\pi\mu 
b^2e^{\varphi_B}\right)d^2 z \right.\nonumber\\
& - & \sum_{n=1}^{N} \left(\eta_n \,
			\frac{1}{4\pi i}\oint_{\partial\Gamma_{n}}
			\varphi_B \,  
			(\frac{dz}{z-z_n}-\frac{d\bar z}{\bar z-\bar z_n})
+\eta_n^2 \log\varepsilon_n^2\right) \nonumber\\
& &\left.+\frac{1}{4\pi i}\oint_{\partial\Gamma_{R}} \varphi_B \, \left(
			\frac{dz}{z} - \frac{d\bar{z}}{\bar{z}}
			\right) + \, \log R^2\right] 
\end{eqnarray}
and
\begin{eqnarray}\label{quantumaction}
S_Q[\varphi_B,\chi] &=& \frac{1}{4\pi}\int_\Gamma\left((\partial_a
\chi)^2 + 4\pi \mu e^{\varphi_B}(e^{2b\chi}-1-2b\chi)\right)d^2z
\nonumber\\
&+& (2+b^2)\ln R^2 +
\frac{1}{4\pi
  i}\oint_{\partial\Gamma_R}\varphi_B\left(\frac{dz}{z}-\frac{d\bar
  z}{\bar z}\right)+ 
\frac{b}{2\pi i}\oint_{\partial\Gamma_R}\chi
\left(\frac{dz}{z}-\frac{d\bar z}{\bar z}\right). 
\end{eqnarray}
The terms in the second row arise from having chosen $Q=1/b+b$.
With regard to the classical action (\ref{classicalaction}) we recall
the Picard inequalities
\begin{equation}\label{picardinequalities}
\left\{ 
\begin{array}{l}
2\eta_n < 1  \\
\sum_{n=1}^{N} \eta_n > 1 
\end{array}
\right.
\end{equation}
which are implied the first by the finiteness of the area and the
second by the Gauss-Bonnet theorem.

\noindent
From the classical action (\ref{classicalaction}) one derives
the equation of motion
\begin{equation}\label{classicalequation}
-\Delta \varphi + 8\pi\mu b^2 \, e^{\varphi} = 8\pi \sum_{n=1}^{N}
 \eta_n \delta^2(z-z_n)
\end{equation}
whose solution is obtained \cite{BG} in terms of two independent
solutions of the fuchsian equation
\begin{equation}\label{eq:fuchsian}
y''(z) + \left( \sum_{n=1}^N \frac{1-\lambda_n^2}{4(z-z_n)^2} +
\frac{\beta_n}{2(z-z_n)} \right) \, y(z) = 0.
\end{equation}
Here in addition to the parameters $\lambda_n = 1-2\eta_n$ related to
the charges, also the Poincar\'e accessory parameters $\beta_n$
appear. These are in principle fixed by the monodromy requirements on
the field $\varphi$. It can be shown that these accessory parameters
must obey three constraints
\begin{equation}\label{fuchsrelations}
\left\{
\begin{array}{l}
\sum_{n=1}^N \beta_n = 0 \\
\sum_{n=1}^N \left[ 2\beta_n z_n + (1-\lambda_n^2) \right] = 0 \\
\sum_{n=1}^N \left[ \beta^{}_n z_n^2 + z_n (1-\lambda_n^2) \right] =
0~.
\end{array}
\right.
\end{equation}
In the case of only three singularities these constraints fix
completely the form of the $\beta_n$, which we report here as we shall
need them later
\begin{equation}\label{eq:accessory-parameters-3-sources}
\left\{
\begin{array}{l}
\beta_1 = \frac{\displaystyle \lambda_1^2 + \lambda_2^2 - \lambda_3^2
	  - 1}{\displaystyle 2(z_1 - z_2)} + \frac{\displaystyle
	  \lambda_1^2 - \lambda_2^2 + \lambda_3^2 - 1}{\displaystyle
	  2(z_1 - z_3)} \\ \beta_2 = \frac{\displaystyle -\lambda_1^2
	  + \lambda_2^2 + \lambda_3^2 - 1}{\displaystyle 2(z_2 - z_3)}
	  -\frac{\displaystyle \lambda_1^2 + \lambda_2^2 - \lambda_3^2
	  - 1}{\displaystyle 2(z_1 - z_2)} \\ \beta_3 =
	  -\frac{\displaystyle -\lambda_1^2 + \lambda_2^2+ \lambda_3^2
	  - 1}{\displaystyle 2(z_2 - z_3)} - \frac{\displaystyle
	  \lambda_1^2 - \lambda_2^2 + \lambda_3^2 - 1}{\displaystyle
	  2(z_1 - z_3)}~.
\end{array}\right.
\end{equation}
The solution of equation (\ref{classicalequation}) is given by
\begin{equation}\label{eq:conformal-factor}
e^{\varphi_c} = \frac{1}{\pi\mu b^2}\,\frac{|w_{12}|^2}{{\left(y_2
\bar{y_2} - y_1 \bar{y_1} \right)}^2}
\end{equation}
where $w_{12}=y_1 y'_2 - y_1'y_2$ is the constant wronskian and the
two solutions  $y_1$ and $y_2$ of (\ref{eq:fuchsian}) must be
chosen in such a way that their monodromy group is $SU(1,1)$ in order
to ensure that the Liouville field $\varphi(z)$ is one-valued on the
whole complex plane.  In the case of only three singularities the
conformal factor is given in terms of hypergeometric functions
\cite{BG}.

We recall two important relations to which the classical action,
i.e. the action (\ref{classicalaction}) computed on the
solution of the equation of motion (\ref{classicalequation}), is
subject. The first is easily derived from the form of the action and
reads
\begin{equation}\label{Xrelation}
\frac{\partial S_{cl}}{\partial \eta_i} = - X_i
\end{equation}
where $X_i$ is the finite part of the field $\varphi$ at $z_i$
\begin{equation} 
\varphi(z) = -2\eta_i\log|z-z_i|^2 +X_i+ o(|z-z_i|).
\end{equation} 
The second relation is the so called Polyakov relation
\cite{ZT1,CMS1,CMS2,ZT2}
\begin{equation}\label{polyakovrelation}
\frac{\partial S_{cl}}{\partial z_i} = -\frac{\beta_i}{2}
\end{equation}
which directly relate the accessory parameters to the classical
Liouville action.

Using these two relations it is possible to compute the semiclassical
limit of the three-point function, which is related
to the value of the classical action. We first consider the case of
singularities placed in $z_1 = 0, z_2 = 1,z_3 = \infty$. The finite
part of the Liouville field $\varphi$ can be computed starting from
equation (\ref{eq:conformal-factor}) and using the explicit form of
the solutions $y_1, y_2$ \cite{BG}. The result is
\begin{equation}
X_1 = -\log (\pi\mu b^2) - \log
\frac{\gamma(\eta_1+\eta_2+\eta_3-1) \gamma(\eta_1-\eta_2+\eta_3)
\gamma(\eta_1+\eta_2-\eta_3)} {\gamma^2({2\eta_1})
\gamma(\eta_2+\eta_3-\eta_1)}
\end{equation}
and cyclic permutations. Here we have defined as usual $\gamma(x) =
\Gamma(x)/\Gamma(1-x)$. Integrating the differential system
(\ref{Xrelation}) one obtains \cite{ZZsphere}
\begin{eqnarray}
S_{cl}[0,1,\infty;\eta_1,\eta_2,\eta_3] &=& S_0 +
		\left(\eta_1+\eta_2+\eta_3 - \frac{3}{2}\right)\log
		(\pi\mu b^2) + 3F(1) \nonumber\\ & &
		-F(2\eta_1)-F(2\eta_2)-F(2\eta_3) +
		F(\eta_1+\eta_2+\eta_3-1) \nonumber\\ & &
		+F(\eta_3+\eta_2-\eta_1) + F(\eta_2+\eta_1-\eta_3) +
		F(\eta_3+\eta_1-\eta_2)
\end{eqnarray}
where the new function $F$ is given by
\begin{equation}
F(x) = \int_{1/2}^{x} \, \log \gamma(s) \, ds.
\end{equation}
The next step is to compute the $z_n$ dependence of the classical
action, that is the semiclassical conformal dimensions of the
Liouville vertex operators. To this purpose we use the Polyakov
relation (\ref{polyakovrelation}) and the explicit form of the
accessory parameters for three singularities
(\ref{eq:accessory-parameters-3-sources}). Equation
(\ref{polyakovrelation}) can be easily integrated to give
\begin{eqnarray}\label{classicalthreepoint}
S_{cl}[z_1,z_2,z_3;\eta_1,\eta_2\eta_3] & = &
			(\delta_1+\delta_2-\delta_3)\log|z_1-z_2|^2  
			+ (\delta_2+\delta_3-\delta_1)\log|z_2-z_3|^2
			\nonumber\\ & &
			(\delta_3+\delta_1-\delta_2)\log|z_3-z_1|^2 +
			S_{cl}[0,1,\infty;\eta_1,\eta_2,\eta_3]
\end{eqnarray}
where $\delta_i = \eta_i ( 1 - \eta_i)$. 

\noindent
In conclusion we have the following expression for the semiclassical
limit of the three-point function
\begin{eqnarray}\label{3pointsemiclassical}
\left< V_{\alpha_1}(z_1) \, V_{\alpha_2}(z_2) \, V_{\alpha_3}(z_3)
\right>_{sc} = C_{sc}(\eta_1,\eta_2,\eta_3) \;
|z_1-z_2|^{-2(\Delta^{sc}_1+\Delta^{sc}_2-\Delta^{sc}_3)} \nonumber\\
|z_2-z_3|^{-2(\Delta^{sc}_2+\Delta^{sc}_3-\Delta^{sc}_1)} \;
|z_3-z_1|^{-2(\Delta^{sc}_3+\Delta^{sc}_1-\Delta^{sc}_2)}
\end{eqnarray}
where
\begin{eqnarray}
&& C_{sc}(\eta_1,\eta_2,\eta_3) = \exp \left( - \frac{1}{b^2}
S_{cl}[0,1,\infty,\eta_1,\eta_2,\eta_3]\right)
				\label{eq:costante-3-punti-semiclassica} 
				\nonumber\\
&& \Delta^{sc}_i = \alpha_i\left(\frac{1}{b} - \alpha_i\right).
\label{eq:dimensione-semiclassica}
\end{eqnarray}
From eq.(\ref{classicalequation}) we can obtain the area of the surface
\begin{equation}\label{area}
A=\int e^{\varphi_c}~d^2z =\frac{1}{\mu b^2}\left(\sum_j\eta_j -1\right).
\end{equation}

\section{The semiclassical four point function}

In this section we shall determine the classical action in presence of
three finite singularities and a fourth infinitesimal; such a calculation
gives the semiclassical four point function for vertices with
three finite charges and the fourth small. As a byproduct we shall
derive in Sect.5 the exact Green function on the sphere with three
arbitrary singularities satisfying the Picard bounds.

The conformal factor with three arbitrary conical defects can be
computed in terms of hypergeometric functions and is well known
\cite{BG}. The possibility of performing such a calculation is related
to the fact that for three singularities the Fuchs relations
(\ref{fuchsrelations}) are 
sufficient for determining the accessory parameters $\beta_j$.  The
procedure we shall use in presence of a fourth weak singularity is to
solve perturbatively the fuchsian equation associated to the Liouville
equation leaving the fourth small accessory parameter $\beta_4$ free,
and then determine it by imposing the monodromy condition on the
conformal factor.

Given four singularities, by means of an $SL(2,C)$ transformation we
can take  three of them in $0,1,\infty$. The position of the fourth will
be called $t$ and the coefficient $Q$ in the fuchsian equation
becomes
\begin{equation}
Q(z) = \frac{1-\lambda_1^2}{4z^2} + \frac{1-\lambda_2^2}{4(z-1)^2} +
	\frac{1-\lambda_4^2}{4(z-t)^2} + \frac{\beta_1}{2z} +
	\frac{\beta_2}{2(z-1)} + \frac{\beta_4}{2(z-t)}~.
\end{equation}
We notice that $1-\lambda_i = 2\eta_i$ and in the case of three
singularities the Picard inequalities (\ref{picardinequalities})
impose $0<\eta_i<1/2$.   
In presence of the fourth singularity the Fuchs relations 
\begin{eqnarray}
\beta_1 & = &
\frac{2+2(t-1)\beta_4-\lambda_1^2-\lambda_2^2-\lambda_4^2+\lambda_3^2}{2}
\nonumber\\ 
\beta_2 & = &
-\frac{2+ 2
  t\beta_4-\lambda_1^2-\lambda_2^2-\lambda_4^2+\lambda_3^2}{2}
\end{eqnarray}
are not sufficient to determine the
$\beta$'s.
For the source in $t$ of infinitesimal strength we shall write
$\lambda_4 = 1-2\varepsilon$ and $\beta_4 = \varepsilon \beta$ and our
aim will be to determine $\beta$. We have
\begin{eqnarray}
\beta_1 & = & \frac{1-\lambda_1^2-\lambda_2^2+\lambda_3^2}{2} +
\varepsilon \, \left[ (t-1)\beta +2\right] + O(\varepsilon^2) \nonumber\\
\beta_2 & = & -\frac{1-\lambda_1^2-\lambda_2^2+\lambda_3^2}{2} -
\varepsilon \, \left[ 2+t\beta \right] + O(\varepsilon^2)
\end{eqnarray}
and we write
\begin{equation}
Q(z) = Q_0(z) + \varepsilon \, q(z)
\end{equation}
where $Q_0(z)$ stays for the coefficient of the three singularity
problem, while $q(z)$ is the perturbation
\begin{equation}\label{eq:q}
q(z) = \frac{1}{2}\left[ \frac{(t-1)\beta+2}{z} - \frac{2+t\beta}{z-1}
+ \frac{\beta}{z-t} + \frac{2}{(z-t)^2} \right]
\end{equation}
After writing $y = y_0 +\delta y$, being $y_0$ a solution of the
unperturbed equation, we have to first order in $\varepsilon$ the
inhomogeneous equation
\begin{equation}\label{inhomogeneous}
(\delta y)'' + Q\, \delta y = -q \, y_0~.
\end{equation}
Such an equation can be solved by a well known method \cite{ince} and
in our case we have
\begin{equation}\label{eq:soluzione-pertubativa}
\delta y_i = -\frac{1}{w_{12}} \int_{z_0}^{z} dx \; \left[ y_1(x) y_2(z) -
y_1(z) y_2(x) \right] \, q(x) y_i(x).
\end{equation}
being $w_{12}= y_1 \,y_2'- y_1'\,y_2$ the constant
wronskian and $z_0$ an
arbitrary base point in the complex plane.  
It will be useful to define the following integrals
\begin{equation}\label{eq:integrali}
I_{ij}(z) \equiv \int_{z_0}^z dx \; y_i(x) y_j(x) \, q(x)
\end{equation}
and in term of them we have the following two independent solutions of
the perturbed problem
\begin{eqnarray}\label{perturbedsolution}
Y_1(z) & = & \left[ 1+\varepsilon \frac{I_{12}(z)}{w_{12}}\right]\, y_1(z)
- \varepsilon\frac{I_{11}(z)}{w_{12}}\, y_2(z)
\label{eq:soluzioni-Y1} \nonumber\\ 
Y_2(z) & = & \varepsilon \frac{I_{22}(z)}{w_{12}}\, y_1(z) + \left[ 1 -
\varepsilon \frac{I_{12}(z)}{w_{12}}\right]\,
y_2(z)~. \label{eq:soluzioni-Y2} 
\end{eqnarray}
We must now compute the monodromy matrices around $0,1,t$ and impose
on them the $SU(1,1)$ nature. This will determine uniquely the
parameter $\beta$ and thus the perturbed conformal factor.
The calculation is given in Appendix A. We find
\begin{equation}\label{beta}
\beta = -4\frac{\kappa \, \bar{y}_1 y_1' - \bar{y}_2 y_2'}{\kappa \,
\bar{y}_1 y_1 - \bar{y}_2 y_2}
\end{equation}
being $\kappa= |k_0|^4$ with $k_0$ the parameter which appears in the
three-singularity conformal factor
\begin{equation}
e^{2b\phi_c^0}= \frac{1}{\pi\mu b^2} \,
\frac{{w_{12}}^2}{({|k_0|^2y_1 \bar{y}_1 - |k_0|^{-2} y_2 \bar{y}_2
)}^2} 
\end{equation}
and
\begin{equation}\label{eq:soluzione-fattore-conforme}
e^{2b\phi_c}=e^{\varphi_c} = \frac{1}{\pi\mu b^2} \,
\frac{{w_{12}}^2}{{\left(Z_1 \bar{Z}_1 - Z_2
\bar{Z}_2\right)}^2}
\end{equation}
with 
\begin{eqnarray}
Z_1(z) & = & k_0 \, \left[ \left( 1 + \varepsilon \frac{I_{12}(z) +
    h}{w_{12}}\right)\, y_1(z) - \varepsilon
    \frac{I_{11}(z)}{w_{12}}\, y_2(z) 
  \right] \label{eq:soluzione-monodroma-1} \nonumber\\ Z_2(z) & = &
\frac{1}{k_0} \, \left[ \varepsilon \frac{I_{22}(z)}{w_{12}}\, y_1(z) +
  \left( 1 - \varepsilon \frac{I_{12}(z) + h}{w_{12}}\right)\, y_2(z)
  \right] \label{eq:soluzione-monodroma-2}
\end{eqnarray}
where the $h$ is also given in Appendix A.
The functions $Z_1,Z_2$ have $SU(1,1)$ monodromies around all
singularities and as such determine a globally monodromic conformal
factor satisfying the Liouville equation.
We can now compute the conformal factor in presence of our four
sources to first order in $\varepsilon$ 
\begin{eqnarray}\label{defchi}
e^{\varphi_c} & = & e^{\varphi_c^0}\left\{ 1 -\eps\, \frac{2}{w_{12}
					\left(\kappa y_1 \bar{y}_1 -
					y_2 \bar{y}_2\right)}
					\right. \nonumber \\ & &
					\left. \qquad\qquad \left[
					\left(\kappa y_1 \bar{y}_1 +
					y_2 \bar{y}_2 \right) \left(
					I_{12} + \bar{I}_{12} + h +
					\bar{h}\right)
					\right. \right. \nonumber \\ &
					& \qquad\qquad \left.\left.-
					y_1 \bar{y}_2 \left( I_{22} +
					\kappa \bar{I}_{11} \right) -
					\bar{y}_1 y_2 \left(
					\bar{I}_{22} + \kappa I_{11}
					\right) \right] +
					O(\varepsilon^2)\right. \! 
					\bigg\}\nonumber\\
					& \equiv & e^{\varphi_c^0} (1 +
					\eps \, \chi +O(\varepsilon^2)).
\end{eqnarray}
Thus
\begin{equation}\label{eq:nuovo-phi}
\varphi_c(z,\bar{z}) = \varphi_c^0(z,\bar{z}) + \eps
\chi(z,\bar{z})+O(\varepsilon^2). 
\end{equation}
The perturbation $\chi$ has a singularity in $t$
\begin{equation}\label{eq:singolarita-chi}
\chi(z,\bar{z}) \sim -4 \, \log|z-t| + c(t) +o(z-t) ~~\mbox{ for}~ z
\rightarrow t
\end{equation}
and in Sect.5 it is proven that $\chi$ is regular in $0,1,\infty$.
Eq.(\ref{beta}) gives the value of $\beta_4$ to first order
$\beta_4=\varepsilon \beta$. Recalling the expression of the
unperturbed conformal factor $e^{\varphi_c^0}$ with only three sources
we have
\begin{equation}\label{betaderivvarphi}
\beta_4 = -4\eps\, {\left. e^{\varphi_c^0/2}\,\partial_z\,
	e^{-\varphi_c^0/2} \right|}_{z=t} = 2\eps \, {\left. \partial_z
	\varphi_c^0(z) \right|}_{z=t}~.
\end{equation}
We can exploit such a result and Polyakov relation to compute to order
$\varepsilon$ the classical action for the new solution
\begin{equation}
\frac{\partial S_{cl}[\eta_1,\eta_2,\eta_3,\eps]}{\partial t} =
-\frac{\beta_4}{2} = -\eps \frac{\partial \varphi_c^0}{\partial t}
\end{equation}
and thus
\begin{equation}\label{eq:azione-classica-4-sorgenti}
S_{cl}[0,1,\infty,t; \eta_1, \eta_2, \eta_3, \eta_4] =
  S_{cl}[0,1,\infty; \eta_1,\eta_2, \eta_3] -\eta_4\, \varphi_c^0(t) +
\eta_4 f(\eta_1,\eta_2,\eta_3)+O(\eta_4^2).
\end{equation}
We exploit now eq.(\ref{Xrelation}) 
\begin{eqnarray}
& &-\frac{\partial
    S_{cl}[0,1,\infty,t,\eta_1,\eta_2,\eta_3,\eta_4]}{\partial \eta_4}= 
\varphi_c^0(t) - f(\eta_1\eta_2,\eta_3) + O(\eta_4)= \nonumber \\
& &{\rm Finite}_{z\rightarrow t}\left(\varphi_c^0(z) +\eta_4(
-4\log|z-t| + c(t) 
+ o(z-t)\right) +O(\eta_4^2)= \nonumber \\
& &\varphi_c^0(t) +\eta_4 c(t) +O(\eta_4^2) 
\end{eqnarray}
from which $f(\eta_1,\eta_2,\eta_3)=0$.

\noindent
Thus for the semiclassical four point function with small $\alpha_4$
we have 
\begin{equation}\label{4pointfunctionspecial}
\left< V_{\alpha_1}(0)  V_{\alpha_2}(1)
	V_{\alpha_3}(\infty) V_{\alpha_4}(t)\right>_{sc} =
	\left<V_{\alpha_1}(0) V_{\alpha_2}(1)
	V_{\alpha_3}(\infty)\right>_{sc} ~e^{2\alpha_4 \phi_c^0(t)}.
\end{equation}
It is easily checked that the four point function
(\ref{4pointfunctionspecial}) has the correct transformation
properties with dimensions 
$\alpha_4/b$ for the vertex field  $V_{\alpha_4}(z_4)$ in agreement
with the semiclassical dimensions $\alpha_4(1/b-\alpha_4)$ keeping in
mind that we have been working to first order in $\alpha_4$, and thus
we can write to first order in $\alpha_4$
\begin{equation}\label{4pointfunction}
\left< V_{\alpha_1}(z_1)  V_{\alpha_2}(z_2)
	V_{\alpha_3}(z_3) V_{\alpha_4}(z_4)\right>_{sc} =
	\left<V_{\alpha_1}(z_1) V_{\alpha_2}(z_2)
	V_{\alpha_3}(z_3)\right>_{sc} ~e^{2\alpha_4 \phi_c^0(z_4)}~.
\end{equation}

\section{Generalization to $n$-point functions}

We can generalize some of the results obtained above to $n$ arbitrary
sources and $m$ infinitesimal sources. Let us start from the case in
which we have $n$ finite sources plus one infinitesimal. Let us
suppose to know a pair of solutions $y_1,y_2$ which produce the monodromic
conformal factor with $n$ finite sources and which we know to
exist. The discussion we have been 
performing on the case of three sources which leads to the
inhomogeneous equation (\ref{inhomogeneous}) remains valid also in
this case; the 
only difference is that now we do not know the explicit form of the
unperturbed solutions $y_1,y_2$. Let us suppose the first three finite
sources 
to be in $0,1,\infty$. Imposition of monodromy in $1$ fixes the
parameter $k$ it happens in the case of three finite singularities
plus one infinitesimal. Then the accessory parameter $\beta_t =
\varepsilon\beta$ is again given by eq.(\ref{betaderivvarphi})
\begin{equation}\label{eq:beta-generale}
\beta = 2 {\left. \partial_z \, \varphi_c^0(z) \right|}_{z=t}
\end{equation}
where now $\varphi_c^0$ is the conformal field which solves the
problem in presence of the $n$ finite sources. Thus we have a general
relation between the value of the accessory parameter relative to the
infinitesimal source in $t$ and the conformal factor for the
unperturbed background and thus we can extend the result
(\ref{4pointfunction}) to 
$n$ finite sources plus an infinitesimal one. Finally due to the
additive nature of the perturbation with $m$ infinitesimal sources we
have in this case for the classical action
\begin{equation}
S_{cl}[z_1,.. z_n,t_1,..t_m;\eta_1,..\eta_n,\eps_1,..\eps_m] =
S[z_1,..z_n;\eta_1,..\eta_n] - 
\sum_{j=1}^m \eps_m \, \varphi_c^0(t_j)
\end{equation}
i.e. the $n+m$ semiclassical correlation function has the value
\begin{equation}
\left< V_{\alpha_1}(z_1)\dots V_{\alpha_n}(z_n)\, V_{\gamma_1}(t_1)
\dots V_{\gamma_m}(t_m) \right>_{sc} = \left< V_{\alpha_1}(z_1)\dots
V_{\alpha_n}(z_n)\right>_{sc} \prod_{j=0}^{m} e^{2\gamma_j
\phi_c^0(t_j)}.
\end{equation}
We stress again that the difference between the case of $n$ finite
sources plus $m$ infinitesimal ones and the case of three finite
sources and one infinitesimal is that in the latter
case  we have an explicit form, in terms of quadratures, of the four
point function.

\section{The Green function on the sphere with three singularities}

From the above derived results we can extract the exact Green function
on the sphere in presence of three finite singularities. The equation
for the Green function is
\begin{equation}\label{greenequation}
-\Delta \, g(z,t) + 8\pi\mu b^2 e^{\varphi_B(z)} \, g(z,t) = 2\pi \,
 \delta^2(z-t)
\end{equation}
where $\varphi_B$ is the classical solution in presence of three
finite singularities.
Such a Green function can be computed from the result obtained in
Sect.3. In fact we have found a solution to
\begin{equation}\label{eq:sorgente-infinitesima-2}
- \Delta \varphi + 8\pi\mu b^2 \, e^{\varphi} = 8\pi \sum_{i=1}^3
  \eta_i \, \delta^2(z-z_i) +8\pi\varepsilon \delta^2(z-t)
\end{equation}
for infinitesimal $\varepsilon$ i.e. $\varphi=\varphi_B+
\varepsilon \chi$. Substituting we obtain
\begin{equation}
-\Delta \chi + 8\pi\mu b^2 e^{\varphi_B}\, \chi = 8\pi\delta^2(z-t)
\end{equation}
i.e. we have $\displaystyle{g(z,t) = \frac{\chi}{4}}$. From
eq.(\ref{defchi}) we have 
\begin{eqnarray}\label{greenfunction}
g(z,t) & = & - \frac{1}{2 w_{12} \left[\kappa y_1(z) \bar{y}_1(\bar{z}) -
	y_2(z) \bar{y}_2(\bar{z})\right]} \Big\{ \left[\kappa y_1(z)
	\bar{y}_1(\bar{z}) + y_2(z) \bar{y}_2(\bar{z}) \right] \,
	\cdot \nonumber \\ & & \qquad \qquad \cdot \left[ I_{12}(z,t)
	+ \bar{I}_{12}(\bar{z},\bar{t}) + h(t) +
	\bar{h}(\bar{t})\right] \nonumber \\ & & \qquad\qquad - y_1(z)
	\bar{y}_2(\bar{z}) \left[ I_{22}(z,t) + \kappa
	\bar{I}_{11}(\bar{z},\bar{t}) \right] \nonumber \\ & & \qquad
	\qquad - \bar{y}_1(\bar{z}) y_2(z) \left[
	\bar{I}_{22}(\bar{z},\bar{t}) + \kappa I_{11}(z,t) \right]
	\Big\}.
\end{eqnarray}
It is possible to verify directly that (\ref{greenfunction}) satisfies
eq.(\ref{greenequation}) by using
\begin{equation}
\frac{\partial I_{ij}(z,t)}{\partial z} = y_i(z) y_j(z) q(z)
\end{equation} and the fact that the wronskian $w_{12} = y_1(z)
y_2'(z) - y_1'(z) y_2(z)$ is constant and real. From this it follows that
expression (\ref{greenfunction}) is completely general, i.e. it
applies also for the case of a background given by $n$ finite sources
with $y_i$ solutions of the related fuchsian equation. Again the
difference with the $n=3$ case is that in the latter case we know the
explicit form of $y_i$ which are not known for $n>3$.
Equation (\ref{greenequation}) for the Green function is obviously
invariant under $SL(2,C)$ transformations. Thus we can write the Green
function for a general location of the singularities in terms of the
standard one with singularities in $0,1,\infty$
\begin{equation}\label{eq:invarianza-funzione-green}
g(z_1,z_2,z_3; z,t) = g\left(0,1,\infty;
	\frac{(z-z_1)(z_3-z_2)}{(z_3-z)(z_2-z_1)},
	\frac{(t-z_1)(z_3-z_2)}{(z_3-t)(z_2-z_1)} \right).
\end{equation}
It is possible to check that the explicit from (\ref{greenfunction})
has the correct logarithmic singularity at $t$
\begin{equation}\label{eq:singolarita-punti-coincidenti}
g(z,t) \sim -\frac{1}{2}\log |z-t|^2 + \dots \qquad \mbox{for } z
\rightarrow t.
\end{equation}
In addition $g(z,t)$ is regular on the three finite sources. By
symmetry it is sufficient to check this e.g. at $z=0$. The integrals
$I_{ij}(z)$ vanish for $z=0$ and thus for $z$ near to 
$0$ we have 
\begin{equation}
g(z,t) \simeq -\frac{|z|^{\frac{1-\lambda_1}{2}} (h + \bar{h})} {2 w_{12}
|z|^{\frac{1-\lambda_1}{2}}} = -\frac{h+\bar{h}}{2 w_{12}}.
\end{equation}

\noindent
Such regularity result allows us to derive a simple relation which
will be useful in the following; integrating eq.(\ref{greenequation})
on the whole complex plane excluding a disk of radius $\varepsilon$
around $t$, 
and then letting $\varepsilon$ going to zero we obtain
\begin{equation}\label{eq:integrale-funzione-green}
\int e^{\varphi_B(z)} \, g(z,t) \, d^2z = 2\pi.
\end{equation}
One would expect the Green function $g(z,t)$ to be symmetric in the 
arguments. This is far from evident from the expression
(\ref{greenfunction}). The 
differential operator $D= -\Delta_{LB}+1$ is hermitean in the
background metric $e^{\varphi_B}d^2z$. As a result also its inverse $G
= D^{-1}$ is hermitean $G=G^+$. $G$ is represented by $g(x,t)$ which
is also real and thus we have $g(z,t) = g(t,z)$.

\section{The quantum determinant}

The complete action is given by eqs.(\ref{classicalaction}) and
(\ref{quantumaction}) and the quantum $n$-point function by
\begin{equation} 
\left< V_{\alpha_1}(z_1)  V_{\alpha_2}(z_2) \dots
V_{\alpha_n}(z_n)\right>= e^{-S_{cl}[\phi_B]} \int D[\chi]~ e^{-S_Q}. 
\end{equation}
We recall that $S_{cl}$ is $O(1/b^2)$ while the quantum action is
\begin{eqnarray}\label{quantumaction2}
S_Q[\varphi_B,\chi] = &
&\frac{1}{2\pi}\int_\Gamma\left(\frac{1}{2}(\partial_a 
\chi)^2 + 2\pi \mu e^{\varphi_B}(e^{2b\chi}-1-2b\chi)\right)d^2z
\nonumber\\ 
&+ &\frac{1}{4\pi i}\oint_{\partial\Gamma_R}
\varphi_B(\frac{dz}{z}-\frac{d\bar z}{\bar z})+ 2 \log R^2 \nonumber\\
&+ & \frac{b}{2\pi i}\oint_{\partial\Gamma_R}\chi
(\frac{dz}{z}-\frac{d\bar z}{\bar z}) + b^2\ln R^2 
\end{eqnarray}
where the first integral can be expanded as
\begin{eqnarray}
& &\frac{1}{4\pi}\int_\Gamma\left ((\partial_a
\chi)^2 + 4\pi \mu e^{\varphi_B}(e^{2b\chi}-1-2b\chi)\right)d^2z
=\nonumber\\
& &\frac{1}{4\pi}\int_\Gamma\left((\partial_a
\chi)^2 + 8\pi \mu b^2 e^{\varphi_B}\chi^2 + 8\pi \mu b^2
e^{\varphi_B}(\frac{4b \chi^3}{3!}+\frac{8 b^2\chi^4}{4!}+\dots\right)
d^2z .  
\end{eqnarray}
From now on we shall denote by $\varphi_B$ the classical solution with
three singularities at $z_1,z_2,z_3$ and with charges
$\eta_1,\eta_2,\eta_3$. 
In performing the perturbative expansion in $b$ we have to keep the
$\eta_1,\eta_2,\eta_3$ constant \cite{ZZsphere}. 
The $O(b^0)$ contribution to the three point function is given by
\begin{equation}\label{determinant}
({\rm Det}D)^{-\frac{1}{2}}= \int D[\chi] e^{-\int \chi(z) D
    \chi(z)f(z)d^2z}
\end{equation}
where $f(z) = 8\pi\mu b^2 e^{\varphi_B(z)}$ and 
\begin{equation}
D=\frac{1}{4\pi}(-\Delta_{LB}+1)
\end{equation}
being $\Delta_{LB}=f^{-1} \Delta$ the Laplace-Beltrami operator on the
background $f(z)$. It provides the one loop quantum correction to the
semiclassical result we have been discussing above.

The usual technique for defining the functional determinant is
provided by the $Z$-function procedure
\begin{equation}
Z_D(s) = {\rm Tr}(D^{-s}) = \sum_n \lambda_n^{-s}
\end{equation}
and
\begin{equation}
Z'_D(0) = -\log ({\rm Det}D) =
\gamma_E Z_D(0) + {\rm Finite}_{\varepsilon \rightarrow
  0}\int_\varepsilon^\infty 
\frac{dt}{t} {\rm Tr}(e^{-t D}).
\end{equation}
We notice that the operator $D$ is covariant under $SL(2,C)$
transformations, thus all the eigenvalues are invariant, and this
gives rise to a definition of ${\rm Det}D$ invariant under $SL(2,C)$.
This means that the semiclassical result for the dimensions of the
vertex fields $e^{2\alpha\phi}$
\begin{equation}
\Delta^{sc}(\alpha)
=\alpha(\frac{1}{b}-\alpha)=\frac{1}{b^2}\eta(1-\eta)
\end{equation}
does not receive any $O(b^0)$ correction, as further quantum
corrections which 
start from order $O(b^2)$ cannot alter such a result.
The situation is similar to the one discussed by D'Hoker, Freedman and
Jackiw \cite{DFJ} and the one considered by Takhtajan \cite{takhtajan}
with the use of an  
invariant regularization of the Green function (Hadamard
regularization \cite{hadamard,garabedian}) and $Q=1/b$. As already
mentioned in the introduction such a regularization scheme gives the
cosmological term the weights $(1-b^2,1-b^2)$. In the following we
shall pursue a different regularization scheme.

\noindent
To compute the determinant (\ref{determinant}) we shall use a
variational procedure similar 
to the one employed in the standard heat kernel approach
\cite{alvarez}, with the 
simplifying feature that we shall compute simply  the derivative with
respect to the three 
parameters $\eta_j$. 
We have  
\begin{equation}\label{etaderivative}
\frac{\partial}{\partial \eta_j}\left(\log ({\rm
  Det}D)^{-\frac{1}{2}}\right)=
-2\pi b^2 \int\frac{\partial\varphi_B}{\partial\eta_1}(z) g(z,z)
  e^{\varphi_B(z)} d^2z.
\end{equation}
In the above equation the Green function at coincident points appears
and such a quantity has to be regularized. We have already seen that
the invariant Hadamard regularization gives rise to a theory in which
the cosmological term $e^{2b\phi(z)}$ does not have weights $(1,1)$ and as
such does not give rise to a theory invariant under the whole
(infinite dimensional) conformal group.

\noindent
We shall adopt here the regularization proposed by Zamolodchikov and
Zamolodchikov \cite{ZZps} (ZZ regulator) for  
perturbative calculations on the pseudosphere i.e.
\begin{equation}
g(z,z) = \lim_{z'\rightarrow z}\left(g(z,z') + \log|z-z'|\right).
\end{equation}  
As under an $SL(2,C)$ transformation
\begin{equation}
w=\frac{az+b}{cz+d}
\end{equation}
the Green function is invariant in value
\begin{equation}
g^w(w,w') = g(z,z')
\end{equation}
we have
\begin{equation}
g^w(w,w) = g(z,z) +\log\left|\frac{\partial w}{\partial z}\right| = g(z,z)
+\log\left|\frac{1}{(cz+d)^2}\right|.  
\end{equation}
The above relation will play a major role in all subsequent
developments.

\noindent
Let us consider first the dilatation $w=\lambda z$; we have $g^w(w,w)
= g(z,z)
+\log|\lambda|$ and thus using eq.(\ref{etaderivative}) keeping in mind
that from eq.(\ref{area})
\begin{equation}
2\pi b^2\int\frac{\partial\varphi_B}{\partial\eta_1}(z)
 e^{\varphi_B(z)}d^2z= 2 
\end{equation}
we have
\begin{equation}
\frac{\partial}{\partial \eta_j}\log\left(({\rm Det}D)^{-\frac{1}{2}}
(\lambda z_1,\lambda z_2,\lambda z_3)\right) = 
\frac{\partial}{\partial \eta_j}\log\left(({\rm Det}D)^{-\frac{1}{2}}
(z_1,z_2,z_3)\right) -  2\log|\lambda|\sum_j\eta_j 
\end{equation}
where we evidenced the dependence of the determinant on the position of
the singularities.
We remark that such a result is consistent with the known exact
structure of the three-point function given by
eq.(\ref{3pointsemiclassical}) with 
$\Delta^{sc}_j$ replaced by the quantum anomalous dimensions
\begin{equation}
\Delta_j = \alpha_j (Q -\alpha_j) = \Delta^{sc}_j + b\alpha_j=
\frac{1}{b^2}\eta_j(1-\eta_j) +\eta_j~.
\end{equation} 
Thus what is left  is to extend the above argument to the general
$SL(2,C)$ transformation and to prove that the subsequent perturbative
corrections $O(b^{2n})$ do not alter the $z_j$ dependence of the
three-point function.
We want to express
\begin{equation}
\frac{\partial}{\partial \eta_j}\log\left(({\rm
  Det}D)^{-\frac{1}{2}}(z_1,z_2,z_3)\right)
\end{equation}
in terms of
\begin{equation}
\frac{\partial}{\partial \eta_j}\log\left(({\rm
  Det}D)^{-\frac{1}{2}}(0,1,\infty)\right). 
\end{equation}
The transformation which takes from $z$ to $w$ is given by
\begin{equation}
w=\frac{(z-z_1)(z_3-z_2)}{(z_3-z)(z_2-z_1)}
\end{equation}
and consequently
\begin{equation}
\varphi_B(z)=\varphi_B^w(w)+2\log\left|\frac{(z_3-z_1)
(z_3-z_2)}{(z_3-z)^2(z_2-z_1)}\right|  
\end{equation}
and for the Green function at coincident points
\begin{equation}
g^w(w,w) =g(z,z) +\log\left|\frac{(z_3-z_1)
(z_3-z_2)}{(z_3-z)^2(z_2-z_1)}\right|.
\end{equation}
We obtain
\begin{eqnarray}
& &\frac{\partial}{\partial \eta_1}\log\left(
({\rm Det}D)^{-\frac{1}{2}}(z_1,z_2,z_3)\right)= 
-2\mu b^2\int\frac{\partial\varphi_B}{\partial\eta_1}(z)
 g(z,z)e^{\varphi_B(z)} d^2z=
\nonumber\\
& &-2\mu b^2
\log\left|\frac{z_2-z_1}{(z_3-z_1)(z_3-z_2)}\right|\frac{\partial
  A}{\partial \eta_1}
-2\mu b^2\int\frac{\partial\varphi^w_B}{\partial\eta_1}(w)
 g^w(w,w)e^{\varphi^w_B(w)} d^2w \nonumber\\
& & -2\mu b^2
\int\frac{\partial\varphi_B(z)}{\partial\eta_1}(z) \log|z-z_3|^2
e^{\varphi_B(z)}d^2z = 
\nonumber\\
& & -2\log\left|\frac{z_2-z_1}{(z_3-z_1)(z_3-z_2)}\right|-2\mu b^2\int
\frac{\partial\varphi^w_B}{\partial\eta_1}(w)
g^w(w,w)e^{\varphi^w_B(w)} d^2w 
-\frac{\partial X_3}{\partial\eta_1}+\frac{\partial
  X_\infty}{\partial\eta_1} \nonumber\\
& & -2\log\left|z_3-z_1\right|^2
\end{eqnarray}
being $X_3$ and $X_\infty$ the finite parts of the field $\varphi_B(z)$
at $z_3$ and $\infty$. The last three terms are the result of
performing the integral  
containing $\log|z-z_3|^2$ which can be computed by using the equation
for the field $\varphi_B(z)$.
Thus
\begin{eqnarray}
& &\frac{\partial}{\partial \eta_1}\log\left(({\rm
 Det}D)^{-\frac{1}{2}}(z_1,z_2,z_3)\right)=\nonumber\\
& &-2\mu b^2\int
\frac{\partial\varphi^w_B}{\partial\eta_1}(z)
 g^w(w,w)e^{\varphi^w_B(z)} d^2w 
-2\log\left|\frac{(z_1-z_2)(z_1-z_3)}{(z_2-z_3)}\right|\nonumber\\
& &-\frac{\partial X_3}{\partial\eta_1}+\frac{\partial
X_\infty}{\partial\eta_1} 
\end{eqnarray}
and similarly
\begin{eqnarray}
& &\frac{\partial}{\partial \eta_2}\log\left(({\rm
  Det}D)^{-\frac{1}{2}}(z_1,z_2,z_3)\right)= \nonumber\\
& &-2\mu b^2\int
\frac{\partial\varphi^w_B}{\partial\eta_2}(z)
  g^w(w,w)e^{\varphi^w_B(z)} d^2w 
-2\log\left|\frac{(z_2-z_3)(z_2-z_1)}{(z_3-z_1)}\right|\nonumber\\
& &-\frac{\partial X_3}{\partial\eta_2}+\frac{\partial
X_\infty}{\partial\eta_2}. 
\end{eqnarray}
On the other hand for the derivative with respect to $\eta_3$ we find
\begin{eqnarray}
& &\frac{\partial}{\partial \eta_3}\log\left(({\rm
  Det}D)^{-\frac{1}{2}}(z_1,z_2,z_3)\right)= \nonumber\\
& &-2\mu b^2\int
\frac{\partial\varphi^w_B}{\partial\eta_2}(w)
  g^w(w,w)e^{\varphi^w_B(w)} d^2w 
-2\log\left|\frac{(z_2-z_1)}{(z_3-z_1)(z_3-z_2)}\right|\nonumber\\
& &-\frac{\partial X_3}{\partial\eta_3}+\frac{\partial
X_\infty}{\partial\eta_3}. 
\end{eqnarray}
Recalling now
eqs.(\ref{Xrelation},\ref{classicalthreepoint})
we see that 
\begin{equation}
\frac{\partial X_3}{\partial \eta_1} =- \frac{\partial
S_{cl}[z_1,z_2,z_3,\eta_1,\eta_2,\eta_3]}{\partial \eta_1\partial
 \eta_3} 
\end{equation}
is independent of $z_i$, and similarly for $\displaystyle{\frac{\partial
X_3}{\partial \eta_2}}$ while we have
\begin{equation}
\frac{\partial X_3}{\partial \eta_3} =- \frac{\partial
S_{cl}[z_1,z_2,z_3,\eta_1,\eta_2,\eta_3]}{\partial
  \eta_3^2}-4\log\left|\frac{z_1-z_2}{(z_2-z_3)(z_1-z_3)}\right|. 
\end{equation}
We must now take into account the additional boundary term contribution
in the action (\ref{quantumaction2})
\begin{equation}
\frac{1}{4\pi i}\oint_{\partial\Gamma_R}
\varphi_B\left(\frac{dz}{z}-\frac{d\bar z}{\bar z}\right)+ 2 \log R^2
= X_\infty 
\end{equation}
which has it origin in the fact that $Q$ is not $1/b$ but $1/b+b$.
Summing up we have for $j=1,2,3$, $k=2,3,1$, $l=3,1,2$
\begin{equation}
\frac{\partial}{\partial \eta_j}\log(({\rm Det}D)^{-\frac{1}{2}}
-\frac{\partial 
X_\infty}{\partial \eta_j}=
f_j(\eta_1,\eta_2,\eta_3)
-2\log\left|\frac{(z_j-z_k)(z_j-z_l)}{(z_k-z_l)}\right|. 
\end{equation}
Integration of the above equation gives
\begin{equation}
c(\eta_1,\eta_2\eta_3) - 2(\eta_1+\eta_2-\eta_3)\log|z_1-z_2|
- 2(\eta_2+\eta_3-\eta_1)\log|z_2-z_3|
- 2(\eta_3+\eta_1-\eta_2)\log|z_3-z_1|
\end{equation}
as $O(b^0)$ correction i.e. the one loop correction,
and we have obtained  the three-point function with the correct
quantum dimensions 
$\Delta_j = \eta_j(1-\eta_j)/b^2 + \eta_j$. Thus the
situation is very similar to what happens on the pseudosphere, where
the one loop corrections with the ZZ regulator provide the exact
quantum dimensions. On the pseudosphere is not too difficult to prove
that higher loop corrections do not change the quantum anomalous
dimensions of the vertex fields $e^{2\alpha\phi}$. On the sphere due
to the appearance of the boundary terms the analysis is more
complicated. We shall give here below the explicit calculation proving
that at two loop no change is induced in the dimensions of the
vertex fields.

\section{Two loop calculation}

In this section we shall compute explicitly the two loop corrections
to the three point function. 
The interaction vertices relevant to the two loop calculation are
\begin{equation}
\frac{b}{6\pi} f(z) \chi^3(z),~~~~
\frac{b^2}{12\pi} f(z)\chi^4(z). 
\end{equation}
with $f(z) = 8\pi\mu b^2 e^{\varphi_B(z)}$.
One should not forget the interaction originating
from the boundary term in eq.(\ref{quantumaction2}) with action given by  
\begin{equation}
S_B =\frac{b}{2\pi i}\oint_{\partial\Gamma_R} \chi(z)
\left(\frac{dz}{z}-\frac{d\bar z}{\bar z}\right).
\end{equation}
The relevant graphs are shown in fig.1.
\begin{figure}
\begin{center}
\epsffile{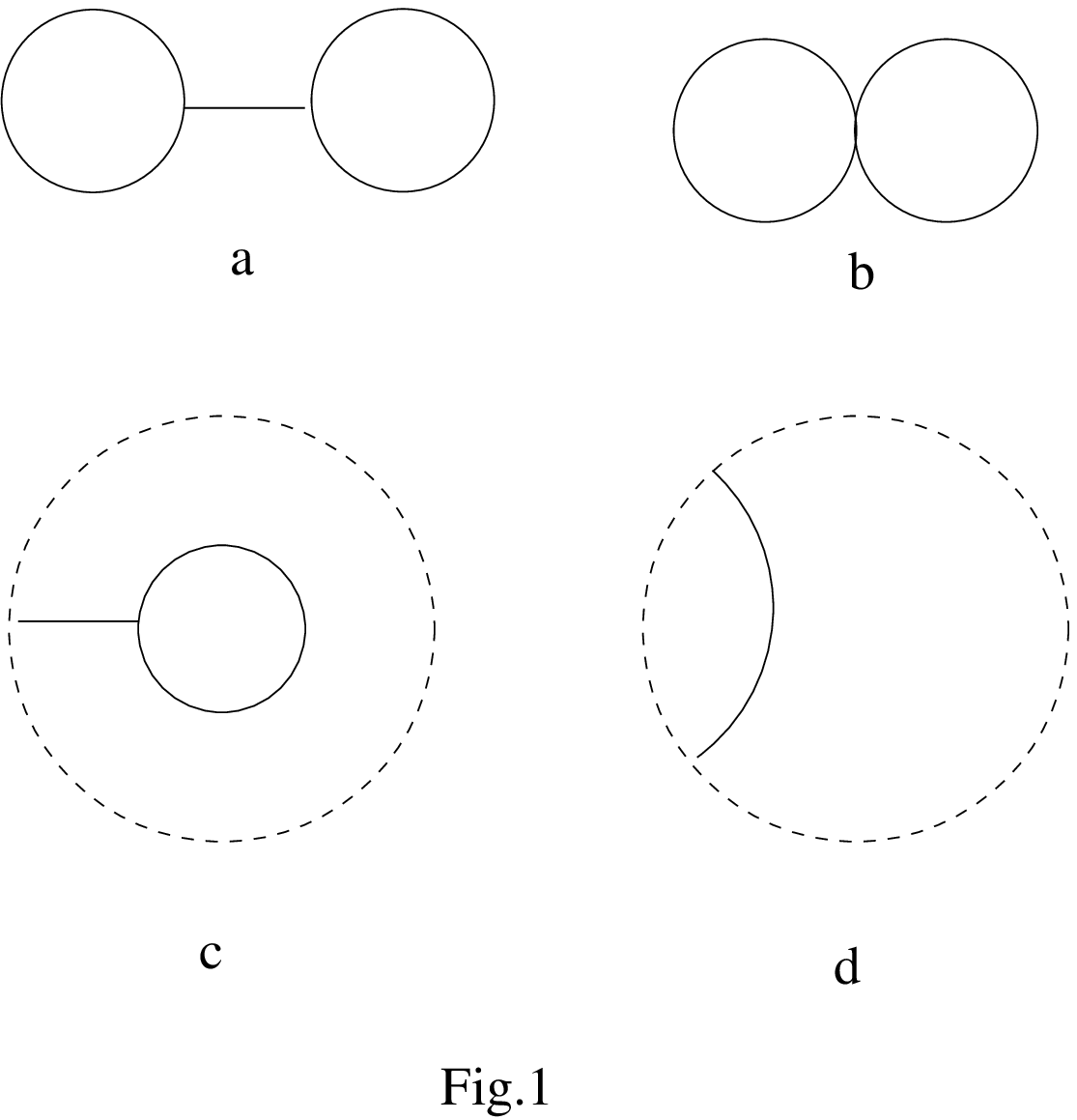}
\end{center}
\end{figure}
The explicit computation gives
\begin{equation}
(a) = \frac{b^2}{8\pi^2}\int g(z,z)f(z)d^2z g(z,z')d^2z' f(z')g(z',z')
\end{equation}
\begin{equation}
(b) = -\frac{b^2}{4\pi}\int g(z,z)f(z)g(z,z)d^2z
\end{equation}
\begin{equation}
(c) = \frac{b^2}{4\pi^2 i}\oint_{\partial\Gamma_R}
  \left(\frac{dz}{z}-\frac{d\bar 
  z}{\bar z}\right) g(z,z')f(z')d^2z'g(z',z')
\end{equation}
and
\begin{equation}
(d) = \frac{b^2}{2(2\pi
  i)^2}\oint_{\partial\Gamma_R}\left(\frac{dz}{z}-\frac{d\bar 
  z}{\bar
  z}\right)g(z,z')\oint_{\partial\Gamma_R}\left(\frac{dz'}{z'}-\frac{d\bar 
  z'}{\bar z'}\right).
\end{equation}
With regard to contribution $(d)$ we notice it can be rewritten by
performing the change of variable $w= -1/z$ as follows 
\begin{equation}\label{dequation}
(d)= b^2 \log R^2 - \frac{b^2}{2\pi}\int_0^{2\pi}
\log(4\sin^2\frac{\phi}{2})~d\phi + 2 b^2 g^w(0,0). 
\end{equation}
We notice that the divergent part in $\log R^2$ is canceled by the
term $-2 b^2\log R$ appearing to order $b^2$ in the expansion of $\exp
(-S)$ and this assures the finiteness of the $O(b^2)$ computation for
$R\rightarrow \infty$.
 
\noindent
We have now to compute the change of the sum $(a)+(b)+(c)+(d)$ under
an $SL(2,C)$ transformation. For doing that it is simpler to prove
separately the invariance under translations, dilatations and
inversions.

\noindent
The invariance under translations is trivial; with regard to
dilatations $w = \lambda z$, $(a)$ goes over to
\begin{eqnarray}
& &(a') = \frac{b^2}{8\pi^2}\int g^w(w,w)f^w(w)d^2w g^w(w,w')d^2w'
  f^w(w')g^w(w',w') = \nonumber\\
& &(a) + \frac{b^2}{4\pi^2 }\log |\lambda|\int f(z)d^2z
  g(z,z')d^2z' f(z')g(z',z')\nonumber\\
& &+\frac{b^2}{8\pi^2}(\log|\lambda|)^2\int f(z)d^2z 
  g(z,z')d^2z' f(z')  
\end{eqnarray}
while $(b)$ becomes
\begin{equation}
(b') = (b)-\frac{b^2}{2\pi}\log|\lambda| \int f(z)g(z,z)d^2z -
\frac{b^2}{4\pi}(\log|\lambda|)^2 \int f(z) d^2z.
\end{equation}
Using the relation
\begin{equation}
\int f(z) ~g(z,z')~  d^2z = 2\pi
\end{equation}
we see that the variation of $(a)+(b)$ vanishes. Similarly one finds
using eq.(\ref{dequation}) that the variation of $(c)+(d)$ vanishes. 

\noindent
We are left to prove the invariance under the inversion $w=-1/z$.
The variation of $(a)$ is computed as usual by performing integrations
by parts
\begin{eqnarray}\label{avariation}
(a') = (a) &-& \frac{ib^2}{2\pi^2}\oint_{\partial\Gamma_R} d\bar z
  \,\log z\bar z \, 
  \partial_{\bar z}g(z,z')\, d^2z' \,f(z') \,g(z',z')\nonumber \\
&-&\frac{b^2}{2\pi}\int d^2z' \,\log z\bar z \,g(z',z')
-\frac{ib^2}{2\pi^2}\oint_{\partial\Gamma_R} \frac{dz}{z} g(z,z')
  f(z') d^2 z' 
  g(z',z')\nonumber\\  
&-&\frac{b^2}{\pi}\int g(0,z') d^2 z' f(z') g(z',z')
+\frac{ib^2}{4 \pi^2}\oint_{\partial\Gamma_R} d\bar z \partial_z
  g(z,z') \log z\bar z f(z') 
   d^2 z' \log z'\bar z'\nonumber\\
&+&\frac{b^2}{4\pi}\int \log z'\bar z' d^2 z' f(z') \log z'\bar z'
+\frac{ib^2}{4\pi^2}\oint_{\partial\Gamma_R} g(z,z') \frac{dz}{z}f(z')
  d^2z' \log z'\bar 
  z'\nonumber \nonumber\\
&+&\frac{b^2}{2\pi}\int g(0,z') f(z') d^2z' \log z'\bar z'
\end{eqnarray}
where the first and the fifth integral vanish in the $R\rightarrow
\infty$ limit due to the appearance of $\partial_{\bar z} g(z,z')$.
On the other hand the variation of $(b)$ is given by
\begin{equation}
(b')=(b) +\frac{b^2}{2\pi}\int \log z\bar z ~f(z) d^2z g(z,z)
-\frac{b^2}{4\pi}\int (\log z\bar z)^2 f(z) d^2z.
\end{equation}
Similarly one computes the variation of $(c)$
\begin{eqnarray}
(c') = (c) &+&\frac{b^2}{\pi}\int g(0,z) d^2z  f(z) g(z,z)
+\frac{b^2}{4\pi^2 i}\oint_{\partial\Gamma_R}
(\frac{dz}{z}-\frac{d\bar z}{\bar z})     
g(z,z') d^2z' f(z') g(z',z')\nonumber\\
& -&\frac{b^2}{\pi}\int g(0,z) d^2z  f(z) \log z\bar z
\end{eqnarray}
and the variation of $(d)$
\begin{eqnarray}
(d') = (d) &-&
\frac{b^2}{8\pi^2}\left(\oint_{\partial\Gamma_{1/R}}(\frac{dz}{z}-
\frac{d\bar
  z}{\bar 
  z}) g(z,z')\oint_{\partial\Gamma_{1/R}}(\frac{dz}{z}-\frac{d\bar
  z}{\bar z})
\right.\nonumber\\ 
&-&\left .\oint_{{\partial\Gamma_R}}(\frac{dz}{z}-\frac{d\bar z}{\bar
  z}) g(z,z')\oint_{{\partial\Gamma_R}}(\frac{dz}{z}-\frac{d\bar
  z}{\bar z})\right).   
\end{eqnarray}
Performing integrations by parts in the third and fourth integrals
appearing in eq.(\ref{avariation}) we find that the variation of
$(a)+(b)+(c)+(d)$ under inversion vanishes and thus we have invariance
under the whole $SL(2,C)$ group. In Appendix B we give the details of
the calculation.

\section{Outlook and conclusions}

In this paper we have provided a functional approach to the quantum
Liouville theory on the sphere, which does not make any appeal to the
canonical quantization. We start from the stable classical
background solution in presence of three finite sources satisfying the
Picard bounds. We compute the semiclassical four point function with
three finite sources and the fourth weak which as a by product
furnishes in terms of quadratures the exact Green function on the
background generated by three finite sources. Several of the results
obtained for the four point functions extend to the case of the
semiclassical vertex functions with $n$ finite charges and $m$
infinitesimal charges. The lowest order quantum correction is provided
by the determinant of the linearized problem on the Liouville
three-source background. The regularization suggested by Zamolodchikov
and Zamolodchikov in the context of the pseudosphere \cite{ZZps} gives
rise to the quantum dimensions found in the hamiltonian approach while
the
invariant (Hadamard) regularization of the Green function fails to
give the cosmological term the weight $(1,1)$. An explicit calculation
shows that the two loop correction do not alter such dimensions. We
expect such a result to hold to all order perturbation theory even if
the presence of the contour terms, which are essential to obtain the
two loop result, makes the calculation not so straightforward as in
the case of the pseudosphere.  The obtained results can be useful to
perform a perturbative analysis of the three-point vertex function
conjectured in \cite{ZZsphere,dornotto} and derived in \cite{teschner}
and of higher point functions.

\section*{Acknowledgments}

One of us (P.M.) is grateful to Domenico Seminara for a useful
discussion.

\section*{Appendix A}

We give here the calculations of the monodromy matrices
around $0,1,\infty,t$ and the procedure to impose on all of them the
$SU(1,1)$ nature.

\noindent
The behavior of the canonical unperturbed solutions around $0$ is
\begin{equation}
y_1(z) \simeq z^{\frac{1-\lambda_1}{2}}, \qquad y_2(z) \simeq
z^{\frac{1+\lambda_1}{2}}~~ {\rm and~}~~ q(z) \simeq
\frac{(t-1)\beta + 2}{2z} + O(1)
\end{equation}
and one can obtain the behavior of the integrals $I_{ij}(z)$ as follows
\begin{eqnarray}
I_{11}(z) & \simeq & I_{11}(0) + \int_0^z dx \;
	x^{1-\lambda_1} \frac{(t-1)\beta+2}{2} x^{-1}\nonumber\\ &
	= & I_{11}(0) + \frac{(t-1)\beta+2}{2(1-\lambda_1)}\,
	z^{1-\lambda_1}
\end{eqnarray}
 \begin{equation}
I_{12}(z) \simeq I_{12}(0) + \frac{(t-1)\beta+2}{2} \; z 
\end{equation}
\begin{equation} 
I_{22}(z) \simeq I_{22}(0) +\frac{(t-1)\beta+2}{2(1+\lambda_1)}\,
z^{1+\lambda_1}
\end{equation}
and of the perturbed solutions
\begin{eqnarray}\label{eq:soluzioni-intorno-0}
Y_1(z) & \simeq & \left[1+\varepsilon\,\frac{I_{12}(0)}{w_{12}}
  \right]z^{\frac{1-\lambda_1}{2}} - \varepsilon\,
\frac{I_{11}(0)}{w_{12}}\,z^{\frac{1+\lambda_1}{2}} \\ Y_2(z) & \simeq &
\varepsilon\, \frac{I_{22}(0)}{w_{12}}\,z^{\frac{1-\lambda_1}{2}} +
\left[1-\varepsilon\,\frac{I_{12}(0)}{w_{12}}\right]
  z^{\frac{1+\lambda_1}{2}}.
\end{eqnarray}
Thus a pair of solution canonical around $0$, i.e. providing around
$0$ a monodromic conformal factor is
\begin{eqnarray}\label{canonicalU}
U_1 & = & \left[1-\varepsilon\,\frac{I_{12}(0)}{w_{12}}\right]\, Y_1 +
	\varepsilon\,\frac{I_{11}(0)}{w_{12}}\, Y_2 \simeq
	z^{\frac{1-\lambda_1}{2}} \nonumber\\ 
U_2 & = &
	-\varepsilon \, \frac{I_{22}(0)}{w_{12}}\, Y_1 +
	\left[1+\varepsilon\,\frac{I_{12}(0)}{w_{12}}\right]\, Y_2 \simeq
	z^{\frac{1+\lambda_1}{2}}\label{canonicalU}.
\end{eqnarray}
Obviously we can apply to the couple of eq.(\ref{canonicalU}) a linear
transformation which leaves unchanged the diagonal form of the
monodromy around $0$ 
\begin{equation}\label{eq:monodromia-0}
M_0 = \left(
\begin{array}{cc}
-e^{-i\pi\lambda_1} & 0 \\
0 & -e^{i\pi\lambda_1}
\end{array}
\right)
\end{equation}
and look for the solution which is monodromic everywhere in the class
of linear combinations. It will be useful to introduce a matrix
notation to deal with such pair of solutions
\begin{eqnarray}\label{eq:matrice-sol-canoniche}
Y & = & \left[1 + \varepsilon\, N(z)\right] \, y \nonumber \\
U & = & (1 - \varepsilon\, P) \, Y = \Lambda\, Y \nonumber \\
N(z) & = & \frac{1}{w_{12}}\, \left(
\begin{array}{cc}
I_{12}(z) & -I_{11}(z) \\
I_{22}(z) & -I_{12}(z)
\end{array}
\right) \nonumber \\
P & = & \frac{1}{w_{12}} \, \left(
\begin{array}{cc}
I_{12}(0) & -I_{11}(0) \\
I_{22}(0) & -I_{12}(0)
\end{array}
\right).
\end{eqnarray}
At this stage we recall the well known result about the monodromy of
the canonical unperturbed solution $y_1(z)$ and $y_2(z)$ around the
point $1$; setting $\zeta =1-z$ we have
\begin{eqnarray}
y_1(z) \simeq A_{11} \, \zeta^{\frac{1-\lambda_2}{2}} + A_{12} \,
\zeta^{\frac{1+\lambda_2}{2}} \nonumber\\ y_2(z) \simeq A_{21} \,
\zeta^{\frac{1-\lambda_2}{2}} + A_{22} \,
\zeta^{\frac{1+\lambda_2}{2}}
\end{eqnarray}
where the matrix $A_{ij}$ is given by
\begin{equation}\label{Amatrix}
A = \left(
\begin{array}{cc}
\frac{\displaystyle \Gamma(c)\Gamma(c-a-b)}{\displaystyle
\Gamma(c-a)\Gamma(c-b)} & \frac{\displaystyle
\Gamma(c)\Gamma(a+b-c)}{\displaystyle \Gamma(a)\Gamma(b)} \\
\frac{\displaystyle \Gamma(2-c)\Gamma(c-a-b)}{\displaystyle
\Gamma(1-a)\Gamma(1-b)} & \frac{\displaystyle
\Gamma(2-c)\Gamma(a+b-c)}{\displaystyle \Gamma(a-c+1)\Gamma(b-c+1)}
\end{array}\right).
\end{equation}
Similarly to what done around $z=0$ we can develop the integrals
$I_{ij}(z)$ around 1 obtaining
\begin{eqnarray}
I_{11}(z) &\simeq& I_{11}(1) - \frac{2+\beta t}{2} \left[
	\frac{A_{11}^2}{1-\lambda_2}\, \zeta^{1-\lambda_2} + 2 A_{11}
	A_{12}\, \zeta + \frac{A_{12}^2}{1+\lambda_2}\,
	\zeta^{1+\lambda_2} \right] \\ 
I_{12}(z) &\simeq& I_{12}(1) -
	\frac{2+ \beta t }{2} \left[
	\frac{A_{11}A_{21}}{1-\lambda_2}\, \zeta^{1-\lambda_2} +
	(A_{11} A_{22} + A_{12}A_{21}) \, \zeta + \right. \nonumber\\ & &
	\qquad \qquad \qquad \qquad \left. +
	\frac{A_{12}A_{22}}{1+\lambda_2}\, \zeta^{1+\lambda_2} \right]
	\\ I_{22}(z) &\simeq& I_{22}(1) - \frac{2+ \beta t }{2} \left[
	\frac{A_{21}^2}{1-\lambda_2}\, \zeta^{1-\lambda_2} + 2 A_{21}
	A_{22}\, \zeta + \frac{A_{22}^2}{1+\lambda_2}\,
	\zeta^{1+\lambda_2} \right].
\end{eqnarray}
Again from eq.(\ref{perturbedsolution}) we have around $z=1$,
$Y = y + \varepsilon\, N\, y$, where
\begin{equation}\label{eq:matrice-n}
N = N(1) = \frac{1}{w_{12}}\left(
\begin{array}{cc}
I_{12}(1) & -I_{11}(1) \\
I_{22}(1) & -I_{12}(1)
\end{array}
\right).
\end{equation}
Thus, if we denote by  $M^{(0)}_1$ the monodromy matrix of the
unperturbed problem around $1$, we have after encircling $1$
\begin{eqnarray}
Y' & = & M^{(0)}_1 y + \varepsilon N M^{(0)}_1 y = M^{(0)}_1
   (1-\varepsilon N) Y + \varepsilon N M^{(0)}_1 Y \nonumber\\ & = & \left(
   M^{(0)}_1 + \varepsilon \left[ N, M^{(0)}_1 \right] \right)\, Y.
\end{eqnarray}
Thus the monodromy matrix around $1$ for the canonical solutions $U$
is
\begin{eqnarray}\label{eq:monodromia-1}
M_1 & = & (1-\varepsilon P)\left( M^{(0)}_1 + \varepsilon \left[ N,
    M^{(0)}_1 \right] \right)(1+\varepsilon P) \nonumber \\ & = &
    M^{(0)}_1 + \varepsilon \left[ N-P, M^{(0)}_1 \right].
\end{eqnarray}
We come to examine the monodromy matrix around the weak singularity
located in $t$. At $t$ the unperturbed solutions are analytic and thus
the only contribution to the monodromy matrix comes from the integrals
$I_{ij}(z)$ 
\begin{equation}\label{eq:variazione-integrali}
I_{ij}(z) \rightarrow I_{ij}(z) + \oint_t \, y_i(x) y_j(x) q(x) \, dx
\end{equation}
to obtain
\begin{eqnarray}\label{eq:delta-I}
\delta I_{ij} & = & \frac{1}{2} \, \oint_t \, y_i(x) y_j(x) 
\left[\frac{\beta}{z-t}+\frac{2}{(z-t)^2}\right] \, dx \nonumber \\
& = & i\pi { \Big[ \beta y_i(x) y_j(x) +2 \left({y_i(x)
	      y_j(x)}\right)'\Big]}_{x=t}
\end{eqnarray}
i.e. we obtain for $Y$ the transformation
\begin{eqnarray}
Y_1 \rightarrow Y_1 + \frac{\varepsilon}{w_{12}}\left[ \delta I_{12} \,y_1
- \delta I_{11} \, y_2 \right] \nonumber\\ Y_2 \rightarrow Y_2 +
\frac{\varepsilon}{w_{12}}\left[ \delta I_{22} \,y_1 - \delta I_{12} \,
y_2 \right]
\end{eqnarray}
and thus to order $O(\varepsilon)$ we have  for the monodromy matrix
 of the canonical solutions $U$
\begin{equation}\label{eq:monodromia-t}
M_t = \Lambda M_t(Y) \Lambda^{-1} = \left(
\begin{array}{cc}
1+\varepsilon\frac{\displaystyle \delta I_{12}}{\displaystyle w_{12}} &
-\varepsilon \frac{\displaystyle \delta I_{11}}{\displaystyle w_{12}} \\
\varepsilon \frac{\displaystyle \delta I_{22}}{\displaystyle w_{12}} &
1-\varepsilon\frac{\displaystyle \delta I_{12}}{\displaystyle w_{12}}
\end{array}
\right) =M_t(Y).
\end{equation}
By construction the canonical solutions $U$ have a monodromy matrix
around $0$ which is an element of $SU(1,1)$. The only freedom left is
the conjugation by a matrix of the form
\begin{equation}\label{eq:matrice-K}
K = \left( \begin{array}{cc}
k & 0 \\
0 & 1/k
\end{array} \right).
\end{equation}
We shall start by imposing the monodromy around $t$. We must impose 
\begin{equation}
\tilde{M}_t \equiv K\,M_t\,K^{-1} = \left(
\begin{array}{cc}
1+\varepsilon\frac{\displaystyle \delta I_{12}}{\displaystyle w_{12}} &
-k^2 \, \varepsilon \frac{\displaystyle \delta I_{11}}{\displaystyle
  w_{12}} \\ \frac{1}{k^2} \, \varepsilon \frac{\displaystyle \delta
  I_{22}}{\displaystyle w_{12}} & 1-\varepsilon\frac{\displaystyle \delta
  I_{12}}{\displaystyle w_{12}}
\end{array}
\right)\in SU(1,1)
\end{equation}
i.e. as such a matrix already belong to $SL(2,C)$ it sufficient to
impose
\begin{eqnarray}
(\tilde{M}_t)_{11} & = & \overline{(\tilde{M}_t)_{22}}
  \label{eq:cond1} \nonumber\\
(\tilde{M}_t)_{12} & = & \overline{(\tilde{M}_t)_{21}} \label{eq:cond2}~~.
\end{eqnarray}
The first condition gives
\begin{equation}\label{firstcondition}
\beta\, y_1 y_2 +2 {\left(y_1 y_2\right)}' = \bar{\beta} \, \bar{y}_1
\bar{y}_2 +2 {\left(\bar{y}_1 \bar{y}_2\right)}'
\end{equation}
where all $y_i$ are computed at $t$.
The second condition gives $k$ as a function of $\beta$
\begin{equation}\label{secondcondition}
|k|^4 = -\frac{\overline{\delta I}_{22}}{\delta I_{11}} =
					\frac{\bar{\beta} \bar{y}_2
					\bar{y}_2 +4 \bar{y}_2
					\bar{y}_2'} {\beta y_1 y_1 + 4
					y_1 y_1'}
\end{equation}
which by the way implies that $\delta I_{11}\delta I_{22}$ is real and
negative which can also be written as 
\begin{equation} \label{inequalitykappa}
\left[ \bar{\beta} \bar{y}_1\bar{y}_2 +2 (\bar{y}_1\bar{y}_2)'
\right]^2 - 4w_{12}^2 > 0.
\end{equation}
We shall check after determining $\beta$ such inequality.
We come now to the imposition of the monodromy at $1$. The
zero order value of $K$ which we shall call $K_0$ can be computed from
the three singularity monodromy matrix (\ref{Amatrix}). Thus we shall pose
\begin{equation}
K = \left(
\begin{array}{cc}
1+\varepsilon \frac{\displaystyle h}{\displaystyle w_{12}} & 0 \\
0 & 1-\varepsilon \frac{\displaystyle h}{\displaystyle w_{12}}
\end{array}
\right)\left(
\begin{array}{cc}
k_0 & 0 \\
0 & 1/k_0
\end{array}\right) = (1+ \varepsilon H) \, K_0~~.
\end{equation}
Notice that as we are free to multiply $K$ by an element
$\displaystyle{{\rm diag}(e^{i\alpha},e^{-i\alpha})}$
of $SU(1,1)$, only the real part of the parameter $h$ is meaningful.
The monodromy matrix of the new solutions is given by
\begin{eqnarray}
K \, M_1 \, K^{-1} & = & (1+\varepsilon H) \,K_0 \, \left( M^{(0)}_1 +
				\varepsilon [ N-P, M^{(0)}_1 ] \right)
				K_0^{-1}\, (1-\varepsilon H)
				\nonumber\\ & = & 
				K_0 M^{(0)}_1 K_0^{-1} + \varepsilon
				[H , K_0 M^{(0)}_1 K_0^{-1} ] +
				\varepsilon K_0 [N-P,M^{(0)}_1]
				K_0^{-1} \nonumber\\ & = & D + \varepsilon
				\left[ H + K_0 
		 		(N-P) K_0^{-1} , K_0 M^{(0)}_1
				K_0^{-1}\right]
\end{eqnarray}
where we set  $D \equiv K_0 M^{(0)}_1 K_0^{-1}$, and we must impose
\begin{equation}
\tilde{M}_1 \equiv K M_1 K^{-1} = D + \varepsilon \left[ H + K_0 (N-P)
	K_0^{-1} , D\right] \equiv D 	+ \varepsilon [B,D]
\end{equation}
to be an element of $SU(1,1)$. The request $D\in SU(1,1)$
is equivalent to the three source problem which has
already been solved by fixing
\begin{equation}\label{eq:valore-k0}
|k_0|^4 = \frac{\overline{M_{21}^{(0)}}}{M_{12}^{(0)}} \equiv \kappa.
\end{equation}
We notice that such a value of $|k_0|^4$ is already sufficient to determine
the value of $\beta$. In fact
Eqs.(\ref{firstcondition},\ref{secondcondition}) furnish the system 
\begin{equation}
\left\{\begin{array}{l} \kappa\, \bar{y}_1 (\bar{\beta} \bar{y}_1 +
 4\bar{y}_1') = y_2 (\beta y_2 +4 y_2') \\ \beta y_1 y_2 +2 (y_1 y_2)'
 = \bar{\beta} \bar{y}_1 \bar{y}_2 +2 (\bar{y}_1 \bar{y}_2)'
\end{array}\right.
\end{equation}
which gives 
\begin{equation}
\beta = -2 \frac{- 2|y_2|^2 y_2' +
\kappa \bar{y}_1 (\bar{y}_2 \bar{y}_1'- \bar{y}_1 \bar{y}_2' + y_2
y_1' + y_1 y_2')}{y_2 
(\kappa \bar{y}_1 y_1 -\bar{y_2} y_2)}~.
\end{equation}
Exploiting the fact that the wronskian of the two solution is real
such expression can be simplified to  
\begin{equation}\label{betaappendix}
\beta = -4\frac{\kappa \, \bar{y}_1 y_1' - \bar{y}_2 y_2'}{\kappa \,
\bar{y}_1 y_1 - \bar{y}_2 y_2}~~.
\end{equation}
We are left to determine the parameter $h$. We have $\det D=1$ and
as a consequence also $\det \tilde M_1 
=1+O(\varepsilon^2)$. The explicit form of the matrix $B$ is       
\begin{equation}\label{eq:matrice-B}
B = \left(
\begin{array}{cc}
\frac{\displaystyle h}{\displaystyle w_{12}} + \frac{\displaystyle
I_{12}(1)-I_{12}(0)}{\displaystyle w_{12}} & -k_0^2 \frac{\displaystyle
I_{11}(1)-I_{11}(0)}{\displaystyle w_{12}} \\ \frac{\displaystyle
I_{22}(1)-I_{22}(0)}{\displaystyle k_0^2 w_{12}} & -\frac{\displaystyle
h}{\displaystyle w_{12}} - \frac{\displaystyle
I_{12}(1)-I_{12}(0)}{\displaystyle w_{12}}
\end{array}
\right)
\end{equation}
which as expected is independent of the base point $z_0$. From now on
we shall fix $z_0=0$ so that $I_{ij}(0)=0$. 
Defined $C=[B,D]$ we must impose  $C_{12} = \bar{C}_{21}$ and $C_{11}
= \bar{C}_{22}$ keeping in mind that already $D\in SU(1,1)$.
The first relation gives the condition
\begin{equation}\label{eq:condizione-2}
B_{11} + \bar{B}_{11} - B_{22} - \bar{B}_{22} = \frac{D_{11} -
		 \bar{D}_{11}}{D_{12}} \, (B_{12}-\bar{B}_{21})
\end{equation}
which can be rewritten as
\begin{equation}\label{eq:valore-h}
h + \bar{h} = - I_{12}(1) - \bar{I}_{12}(1) +
		\frac{A_{12}A_{21}+A_{11}A_{22}}{2} \left(
		\frac{I_{11}(1)}{A_{11}A_{12}} +
		\frac{\bar{I}_{22}(1)}{A_{21}A_{22}} \right).
\end{equation}
The second relation can be rewritten as
\begin{equation}
A_{21} A_{22} \, I_{11}(1) - A_{11} A_{12} \, I_{22}(1) \in {R}.
\end{equation}
Due to the solubility of the four source problem as assured by Picard
theorem such a relation has to be satisfied and furnishes a non trivial
relation among the integrals $I_{11}$  and $I_{22}$ containing
hypergeometric functions. We did not find such a relation in the
standard tables but as a check we verified it numerically to $10^{-12}$
precision. 
Thus we have reached the following pair of solutions
\begin{eqnarray}
Z_1(z) & = & k_0 \, \left[ \left( 1 + \varepsilon \frac{I_{12}(z) +
    h}{w_{12}}\right)\, y_1(z) - \varepsilon
    \frac{I_{11}(z)}{w_{12}}\, y_2(z)
  \right] \label{eq:soluzione-monodroma-1} \nonumber\\ Z_2(z) & = &
\frac{1}{k_0} \, \left[ \varepsilon \frac{I_{22}(z)}{w_{12}}\, y_1(z) +
  \left( 1 - \varepsilon \frac{I_{12}(z) + h}{w_{12}}\right)\, y_2(z)
  \right] \label{eq:soluzione-monodroma-2}
\end{eqnarray}
which have $SU(1,1)$ monodromies around all singularities and as such
determine a globally monodromic conformal factor satisfying the
Liouville equation
\begin{equation}\label{eq:soluzione-fattore-conforme}
e^{2b\phi_c}=e^{\varphi_c} = \frac{1}{\pi\mu b^2} \,
\frac{{w_{12}}^2}{{\left(Z_1 \bar{Z}_1 - Z_2
\bar{Z}_2\right)}^2}
\end{equation}
as $Z_1 Z_2' - Z_1' Z_2 =w_{12}= {\rm const }$.
At this stage we can verify the inequality
(\ref{inequalitykappa}). Substituting eq.(\ref{betaappendix}) into
(\ref{inequalitykappa}) we have
\begin{equation}
\left[ \bar{\beta} \bar{y}_1\bar{y}_2 + 2 (\bar{y}_1\bar{y}_2)'
	\right]^2 - 4w_{12}^2 = 16 \frac{\kappa |y_1|^2 |y_2|^2
	w_{12}^2}{(\kappa |y_1|^2-|y_2|^2)^2} > 0.
\end{equation}

\section*{Appendix B}
We give here the details of the calculation which proves the invariance
under inversion of the two-loop contribution.

\noindent
The first integral in eq.(\ref{avariation}) vanishes for $R\rightarrow
\infty$ as  the infinity is a regular point  for the Green function
$g(z,z')$ and as such $\partial_{\bar z}g(z,z') = O(1/|z|^2)$ and the
same holds for the fifth integral.
The $(d') -(d)$ term performing the change of variable $w=-1/z$ can be
rewritten as
\begin{equation}
(d') -(d) = 2b^2\left(g(0,0) - g^w(0,0)\right).
\end{equation}
Then summing all contributions we find
\begin{eqnarray}\label{granddifference}
& &(a')+(b')+(c')+(d') -(a)-(b)-(c)-(d) = 
\frac{b^2 i}{4\pi^2}\oint_{\partial\Gamma_R}\frac{dz}{z}g(z,z')d^2z' 
\log z'\bar z'\nonumber\\
&-&\frac{b^2}{2\pi}\int g(0,z') d^2z' f(z')\log z'\bar z'
+2 b^2\left(g(0,0) - g^w(0,0)\right)\nonumber\\
&=&\frac{b^2}{2\pi}\int g^w(0,w')f^w(w') d^2 w' \log w'\bar w'
-\frac{b^2}{2\pi}\int g(0,z')f(z') d^2 z' \log z'\bar z'\nonumber\\
&+&2 b^2\left(g(0,0) - g^w(0,0,)\right).
\end{eqnarray}
Using the differential equation for $g(0,z')$ and integrating by parts
we find
\begin{equation}
\frac{b^2}{2\pi}\int g(0,z) f(z) d^2 z \log z\bar z = 2 b^2 g(0,0) - 
2b^2 g(0,\infty) 
\end{equation}
and thus we find for eq.(\ref{granddifference}) the value
\begin{equation}
2b^2 g(0,\infty) -2b^2 g^w(0,\infty) =0 
\end{equation}
due to the invariance in value of the Green function $g(z,z') =
g^w(-1/z,-1/z')$ and the symmetry of the Green function.

\end{document}